\documentclass[fleqn,usenatbib]{mnras}

\usepackage{newtxtext,newtxmath}
\usepackage[T1]{fontenc}

\DeclareRobustCommand{\VAN}[3]{#2}
\let\VANthebibliography\thebibliography
\def\thebibliography{\DeclareRobustCommand{\VAN}[3]{##3}\VANthebibliography}

\usepackage{graphicx}	% Including figure files
\usepackage{amsmath}	% Advanced maths commands
\title[Predicting Dust with Machine Learning]{Dust Extinction Measures for $z\sim 8$ Galaxies using Machine Learning on JWST Imaging}

\author[Fu et. al]{
Kwan Lin Kristy Fu,$^{1}$\thanks{E-mail: kwanlinkristy.fu@postgrad.manchester.ac.uk (KTS)}
Christopher J. Conselice,$^{1}$
Leonardo Ferreira,$^{2}$
Thomas Harvey,$^{1}$
Qiao Duan,$^{1}$
\newauthor
Nathan Adams,$^{1}$
Duncan Austin$^{1}$
\\
$^{1}$Jodrell Bank Centre for Astrophysics, University of Manchester, Oxford Road, Manchester, UK \\
$^{2}$ School of Physics and Astronomy, University of Victoria, Victoria, BC, Canada
}

\date{Accepted XXX. Received YYY; in original form ZZZ}

\pubyear{2022}

\begin{document}
\label{firstpage}
\pagerange{\pageref{firstpage}--\pageref{lastpage}}
\maketitle

\begin{abstract}

We present the results of a machine learning study to measure the dust content of galaxies observed with JWST at z > 6 through the use of trained neural networks based on high-resolution IllustrisTNG simulations. Dust is an important unknown in the evolution and observability of distant galaxies and is degenerate with other stellar population features through spectral energy fitting. As such, we develop and test a new SED-independent machine learning method to predict dust attenuation and sSFR of high redshift (z > 6) galaxies. Simulated galaxies were constructed using the IllustrisTNG model, with a variety of dust contents parameterized by E(B-V) and A(V) values, then used to train Convolutional Neural Network (CNN) models using supervised learning through a regression model. We demonstrate that within the context of these simulations, our single and multi-band models are able to predict dust content of distant galaxies to within a 1$\sigma$ dispersion of A(V) $\sim 0.1$. Applied to spectroscopically confirmed z > 6 galaxies from the JADES and CEERS programs, our models predicted attenuation values of A(V) < 0.7 for all systems, with a low average (A(V) = 0.28). Our CNN predictions show larger dust attenuation but lower amounts of star formation compared to SED fitted values. Both results show that distant galaxies with confirmed spectroscopy are not extremely dusty, although this sample is potentially significantly biased. We discuss these issues and present ideas on how to accurately measure dust features at the highest redshifts using a combination of machine learning and SED fitting.

\end{abstract}


%%%%%%%%%%%%%%%%%%%%%%%%%%%%%%%%%%%%%%%%%%%%%%%%%%

%%%%%%%%%%%%%%%%% BODY OF PAPER %%%%%%%%%%%%%%%%%%

\section{Introduction}

Interstellar dust represents potentially a significant unknown in observing early galaxies in optical and near-infrared wavelengths \citep[e.g.,][]{Casey2014, Fudamoto2020}. Due to dust effects, such as extinction and the scattering of light, images of early galaxies are often obscured, and properties such as dust composition \citep{Calzetti1994, Min2006, Spoon2006, Dwek2014}, AGN \citep{Treister2010}, and metallicity \citep{Shivaei2020} can affect spectral energy distribution modelling accuracy \citep[e.g.,][]{Ignas2023}. On the other hand, solid measurements of the dust content or reddening can help build accurate models of galaxies, and are important in our understanding of galaxy evolution. We are now in a position, with ALMA and JWST, to learn more about dust in the early universe and how dust affects the output of light from early galaxies. However, there remains a significant amount we do not yet understand about dust in the early universe, in part due to the difficulties of observing this feature within faint high-redshift galaxies in the absence of rest-frame far-IR light. New approaches and ideas are needed to trace this aspect in the early universe.   

Observations starting from IRAS and COBE in 1980s and 90s and leading into JCMT/SCUBA, SMO, and ALMA, among others, have identified many high-redshift galaxies which are faint or invisible in the optical but bright in the far-IR/sub mm \citep[e.g.,][]{Casey2014, Hodge2020, Dayal2022}. These galaxies are sometimes known as Dusty Star-Forming Galaxies (DSFGs) due to being enshrouded in dust, which obscures observations at optical and UV wavelengths. Observed DSFGs are generally massive, ultraluminous, and have high Star Formation Rates (SFRs) \citep{Hodge2013}. DSFGs include sub-millimeter galaxies (SMGs), systems observed to be the brightest systems at 850 $\mu$m but are incredibly faint in other bandwidths \citep{Blain2002}. High-\textit{z} SMGs are important as they are thought to contribute significantly to the SFR and energy in this time period, where the average SFR is $\sim$ 200-300 and can reach 1000 $M_{\bigodot} \textup{ yr}^{-1}$ \citep[e.g.,][]{Blain2002, Daddi2005}, making up a significant portion of SFR in this time period \citep{Chapman2005}. Large amounts of stellar light heats up the dusty star forming environment which become extremely bright in the infrared. This can impact measurements as high dust temperatures increase the sub-mm observed flux by up to a factor of 10 \citep{Blain2002, Eales1999}, causing an overestimation of the number density of high-\textit{z} galaxies \citep{Hodge2020}.   These are, however, for the most part extreme and relatively rare systems and the effects of dust on more typical galaxies is harder to know for certain.

%However, dust is not just important in galaxies where the energetic output is dominated by it. 
Aside from energetic, high star-forming galaxies, dust is also important in understanding how it affects observations of galaxies, including their measured stellar masses and star formation rates, among other properties. For example, \citet{Fudamoto2017} suggests that there is an evolution in the dust attenuation curve between \textit{z} $\sim$ 3 to 6 that produces observable effects. Remarkably, ALMA shows that even UV bright galaxies, whose light would not normally be thought to be attenuated much by dust, often have their star formation dominated by dust emission at $z \sim 7$ \citep[e.g.,][]{Dayal2022, Inami2022}.  It is thus important to learn more about the dust content at high redshifts to accurately constrain the properties of distant galaxies, such as ongoing and past star formation, metallicity, and with spectral fits of SED properties based on their spectral energy distribution. 

Another reason for this is a full accounting of galaxy formation.  One aspect of this is measuring the star formation history of galaxies. Measurements of cosmic star formation rate density (SFRD) has been shown to reach a peak at \textit{z} $\sim$ 2 \citep{Juneau2005, Madau2014} with SFR dropping off at either ends of the peak, but there may still be significant amounts of star formation at very high redshifts \textit{z} > 8 \citep{Ellis2012, Robertson2015, Duncan2014, Adams2023}.   For example, \citet{Daddi2009} suggests that a significant percent of SMGs are located at \textit{z} > 6, indicating that dust clouds were well enriched before this time period for star formation, showing the importance of SMGs in early cosmic SFRD. Recent observations using ALMA and JWST at redshifts $\geq$ 5 \citep{Blain2002, Coe2012, Oesch2015, Naidu2022, Adams2023, Bowler2023} show that dust may still be an important aspect even back to the very first galaxies, yet it has proven difficult to constrain this.

Other studies utilising the HST \citep{Bouwens2007, Dunlop2011, Wilkins2011, Robertson2015} find significant amounts of dust attenuation and evolution of dust at higher redshifts. ALMA has allowed for a more extensive look into the dust content of these high redshift galaxies. \citet{Laporte2017} measures the physical properties of a \textit{z} = 8.38 galaxy, finding a star formation rate of $\sim$ 20 M$_{\odot}$ year$^{-1}$ and dust extinction value of $A_{V}$ = 0.74. Other ALMA studies of dust content in high redshift galaxies \citep[e.g.,][]{Watson2015, Hodge2020, Casey2019, Bowler2023} find a wide range of dust extinction values, up to $A_{V}$ = 5.64.  Using the REBELS ALMA program \citet{Bowler2023} determined through stacking galaxies that distant galaxies at $z = 4-8$ show significant dust extinction. This wide range of galaxy properties show the need for more data in this regime to understand better how to quantify the amount and impact of dust without requiring deep sub-mm observations, such as with ALMA, which is all but impossible to measure for many very high redshift galaxies. 

There is however some lack of consensus  about the amount of dust-obscured star formation. \citet{Casey2014, Inami2022, Barrufet2023} estimates significant amounts of dust-obscured star formation at \textit{z} > 6, whereas other studies suggest otherwise \citep{Zavala2021, Algera2022}. Observations of high redshift galaxies still presents many difficulties and are constrained by technical limitations. New instruments such as JWST can potentially allow for more information on the dust attenuation evolution in high-redshift environments.  Papers such as \citet{Faisst2024, Ferrara2023, Palla2023, Ziparo2023} investigate and constrain dust attenuation from JWST observations, also showing that high-redshift galaxies from JWST may be bluer than previously thought \citep{Cullen2023, Markov2024}, implying that dust is not a significant presence within these early galaxies.  Limitations and degeneracies in SED fitting techniques also make it difficult to accurately measure dust content without extensive data at mid and far infrared wavelengths. 

We thus turn to machine learning (ML) techniques to determine the amount of dust extinction in high redshift galaxies. Instruments such as JWST promise to deliver higher fidelity images and see further back to earlier galaxies, allowing us to observe and study galaxies in their infancy. In recent years, the development of ML algorithms has made them easier and more adaptive to learn and utilise. ML is potentially capable of using complex physical models to extract physical parameters such as the dust extinction and sSFR of high redshift galaxies. In the context of astrophysics, we have seen the application of ML in uses such as galaxy classification \citep[e.g.,][]{Hocking2017, Cheng2020}, planetary atmospheres \citep{Hayes2020}, light curves \citep{Tarsitano2022}, and quantifying morphology \citep{Tohill2020}. 

This paper uses deep learning to pick out galaxy features that the human eye cannot identify (as opposed to galaxy classification), and uses the power of neural networks to predict properties for high redshift galaxies. We focus our attempts on using features from the morphology of high-redshift galaxies to determine if ML algorithms can take advantage of the detailed images that JWST offers. As galactic structure can depend on observed wavelength \cite{Kuchinski2000,Mager2018}, the difference in the morphology between   JWST wavelengths could provide a way to estimate dust effects and other properties of high-redshift galaxies.  

In this paper we  discuss our use of convolutional neural networks (CNN), a type of machine learning algorithm, to estimate the colour excess and total extinction values of high-redshift galaxies, and our attempts in applying the CNN model onto JWST data. The outline of this paper is as follows: Section \ref{sec:data}   gives an overview into the data and algorithm that we use, while Section \ref{sec:results} shows the results on both simulated as well as high-redshift galaxy candidates from newly released JWST data. Finally, a discussion of the results in Section \ref{sec:disc}, possible improvements and future direction in the conclusion. Throughout this paper we use the Planck cosmology. 

\section{Data and Methods}
\label{sec:data}

\subsection{Simulated Data}

\begin{figure}
    \centering
    \includegraphics[width=\linewidth]{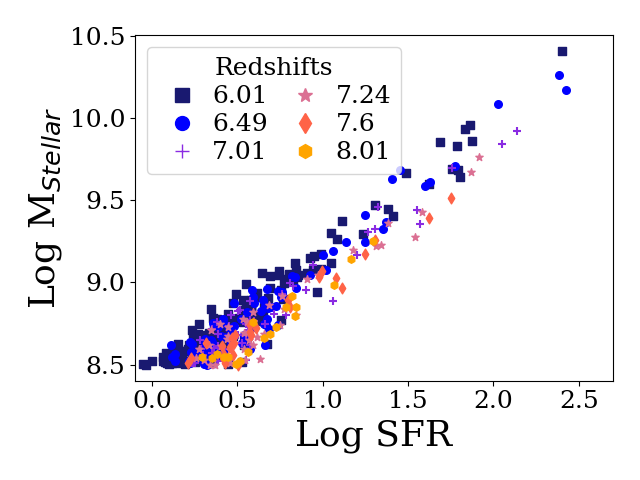}
    \caption{\textbf{SFR and M$_{Stellar}$ of our simulated galaxies : } Values of Log~SFR against Log~M$_{Stellar}$ of the Illustris TNG simulated galaxies we use within this paper. 
 The colour of the points represent the value of the redshift for each galaxy. Simulated galaxies we use from Illustris are found at redshifts between 6 < \textit{z} < 8. Redshifts values are discrete and are noted in the legend with different colors and markers. There is a linear relationship between these two properties. At the same SFR, younger galaxies appear to have a larger LogM$_{Stellar}$ value than the older galaxies.}
    \label{fig:mstar}
\end{figure}

\begin{figure*}
    \centering
        \begin{tabular}{c c c c c c c}
            ID & F277W & F356W & F444W & F560W & F770W & F1000W \\
            8\_2405 & \includegraphics[width=.125\linewidth]{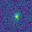} & \includegraphics[width=.125\linewidth]{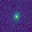} & \includegraphics[width=.125\linewidth]{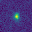} & \includegraphics[width=.125\linewidth]{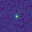} & \includegraphics[width=.125\linewidth]{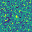} & \includegraphics[width=.125\linewidth]{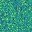} \\
            9\_3842 & \includegraphics[width=.125\linewidth]{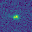} & \includegraphics[width=.125\linewidth]{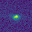} & \includegraphics[width=.125\linewidth]{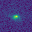} & \includegraphics[width=.125\linewidth]{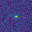} & \includegraphics[width=.125\linewidth]{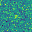} & \includegraphics[width=.125\linewidth]{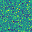} \\
            10\_6345 & \includegraphics[width=.125\linewidth]{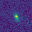} & \includegraphics[width=.125\linewidth]{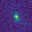} & \includegraphics[width=.125\linewidth]{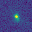} & \includegraphics[width=.125\linewidth]{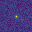} & \includegraphics[width=.125\linewidth]{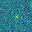} & \includegraphics[width=.125\linewidth]{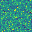} \\
            11\_384 & \includegraphics[width=.125\linewidth]{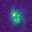} & \includegraphics[width=.125\linewidth]{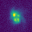} & \includegraphics[width=.125\linewidth]{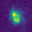} & \includegraphics[width=.125\linewidth]{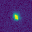} & \includegraphics[width=.125\linewidth]{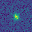} & \includegraphics[width=.125\linewidth]{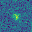} \\
            12\_14814 & \includegraphics[width=.125\linewidth]{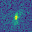} & \includegraphics[width=.125\linewidth]{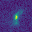} & \includegraphics[width=.125\linewidth]{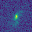} & \includegraphics[width=.125\linewidth]{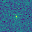} & \includegraphics[width=.125\linewidth]{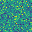} & \includegraphics[width=.125\linewidth]{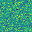} \\
            13\_3717 & \includegraphics[width=.125\linewidth]{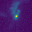} & \includegraphics[width=.125\linewidth]{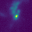} & \includegraphics[width=.125\linewidth]{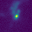} & \includegraphics[width=.125\linewidth]{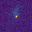} & \includegraphics[width=.125\linewidth]{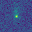} & \includegraphics[width=.125\linewidth]{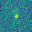} \\
        \end{tabular}
        \caption[Simulated Images]{\textbf{Simulated Images}: A selection of IllustrisTNG images used to train the CNNs. Each galaxy is identified by its ID and is simulated in both NIRcam and MIRI wavelengths. Images here are shown with a dust to metal ratio of 0.1 and in the same orientation. MIRI band images show much more noise and can make it difficult to identify features from galaxies as compared to NIRcam bands.}
        \label{fig:samp}
\end{figure*}

The simulated images we use in this paper are generated from the IllustrisTNG TNG50-1 simulation \citet{Vogelsberger2020,  Marinacci2018, Naiman2018, Nelson2018, Nelson2019, Pillepich2018, Springel2018} forward modelled with the SKIRT radiative transfer code. This ensures that we are simulating how these galaxies would appear when observed in different JWST filters with the Near-Infrared Camera (NIRCam) and Mid-Infrared Instrument (MIRI). 

We generate this mock imaging dataset using the methods outlined in \citet{Ferreira2022}. Here we give a brief description of the process. First, all the galaxies between redshifts \textit{z} $\sim$ 6 to 8 with stellar masses above $M_* > 10^8 M_\odot$ are selected from TNG50-1. From these, all the stellar particles and gas cells in a region centered in the galaxy extending out to a $60$ kpc field of view are used. 

Second, as dust is not directly embedded in the simulations, we model it from the dense cold gas assuming a dust-to-metal ratio \citep{Camps2016, Rodriguez2019}. Instead of using a fixed dust-to-metal ratio as in \citet{Ferreira2022}, we generate different realizations of the same galaxies with dust-to-metal ratios varying from $0$ to $1$ in $0.1$ bins. SKIRT uses the dust distribution to generate a dust screen that will impact the final SED and morphology of the galaxy with the associated reddening and extinction based on the optical depth of the dust. 

Third, the stellar particles, the gas cells and dust distribution are given as input to SKIRT. SKIRT is set-up to generate IFU datacubes in four different orientations aligned with the simulation axes (xy, xz, yz, and an octant view) ranging from the rest-frame UV to the far infrared.  Finally, these datacubes are processed into broadband images based on the filter response curves for NIRCam and MIRI at the observed frame of the source redshift, as well as an integrated spectra for each orientation.

All of these simulations were run with standard cosmology parameters from the {\it Planck} \cite{Planck2016}. Smaller image sizes help reduce memory requirements and increase computational speed, so images were cropped to 32 $\times$ 32 sized arrays centered on the target galaxy before training. Each image is normalized to prevent effects of differing fluxes in different galaxies. Properties of the simulated galaxies include color indices, dust attenuation, dust mass ($M_{dust}$), gas mass ($M_{gas}$), and star formation rate (SFR). A sample of the simulated images we use within this paper are shown in Figure \ref{fig:samp} within six JWST broad-band filters. The simulated images show both NIRcam and MIRI wavelengths, but only the NIRcam (F277W, F356W, F44W) bands are used in this project. This reduced the number of parameters in the machine, helping speed up training time and reducing the memory required. Figure \ref{fig:mstar} shows the properties LogSFR and LogM$_{Stellar}$ of the simulated galaxies with the colour showing the redshift of each system. There appears to be a linear relationship between the two values, consistent with a main sequence of star formation.  From this we can see that our systems are dominated by star formation and no systems have becoming 'passive' or quenched since their formation. 

\subsection{Dust Formalism}

Our primary goal in this project is to retrieve the physical parameters: dust extinction ($A_{V}$), colour excess ($E(B-V)$) and specific star formation rate (sSFR). Using dust extinction and colour excess, we calculate the $R$-value by using the relation :

\begin{equation}
    R_{V} = \frac{A_{V}}{E(B-V)}
    \label{eq:rv}
\end{equation}

\begin{figure}
    \centering
    \includegraphics[width=\linewidth]{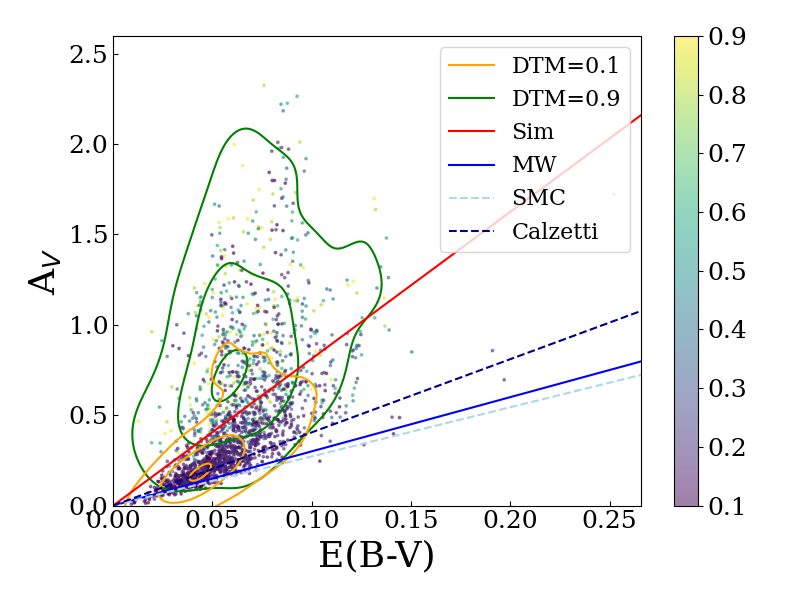}
    \caption[R Values]{\textbf{\textit{ Plot of R} values}: Scatterplot of the simulated \textit{A}$_{V}$ and $E(B-V)$ values from the Illustris images, used to calculated a set of \textit{R} values for training by using Eq \ref{eq:rv}. In this set of simulated data, the red line shows the median value of the calculated R-value and can be compared to the three blue lines representing the average R-values from MW, SMC and Calzetti respectively \citep{Calzetti2001}. The points in the plot are color-mapped to the dust to metal ratio (DTM) value used to generate the images. Images with high DTM ratios have larger R-values, shown by using density contours over the points. The orange contour encircles a proportion of the dust values which have DTM=0.1, while the green contour shows the density of points using DTM=0.9.}
    \label{fig:rvalue}
\end{figure}

\noindent This equation can be rewritten to calculate the dust extinction at any wavelength $\lambda$ as given by:

\begin{equation}
    {A_{\lambda} = k(\lambda) \times E(B-V) = \frac{k(\lambda) A_{V}}{R_{V}}},
    \label{eq:dust}
\end{equation}

%http://www.bo.astro.it/~micol/Hyperz/old_public_v1/hyperz_manual1/node10.html

\noindent where the value $k(\lambda)$ is the reddening curve. The value of R$_{V}$ is thought to be more or less constant within single galaxies, with values of R$_{V} = 3.1$ for the Milky Way, R$_{V} \sim 2.7$ for the SMC, and for the Calzetti law R$_{V} = 4.05$ \citep{Calzetti1994, Fitzpatrick1999}. Figure \ref{fig:rvalue} shows the colour excess against the total extinction, whereby both values used to calculate the R-value. In this simulated dataset, the median of the R-value for the simulated galaxies is $\sim$ 7, comparatively higher than the median R-value of galaxies such as the Milky Way, which has a R-value of around 3 \citep{Fitzpatrick1999}. Since the dust-to-metal (DTM) ratios that we use to generate the simulated images varies from 0 to 1, we expect this to have an impact on the extinction values. To breakdown the effect of various DTM values to the high R-values, we use contour plots in Figure \ref{fig:rvalue} to show the scatter of low DTM and high DTM. The contour plot shows the density of the scatter at 10, 50 and 95\% of the points. The orange contour around DTM=0.1 values shows that the majority of R-values of the DTM=0.1 points has a similar R-value as \citet{Calzetti1994}. This can be contrasted to the green contour around the values of DTM=0.9, which shows a much larger scatter and deviates away from the R-values we would expect. The number of galaxies with DTM=0.1 to DTM=0.9 is roughly 9:1, thus most systems are within the range of the Calzetti dust law.

\begin{figure}
    \centering
    \begin{tabular}{c c c}
        F277W & F356W & F444W \\
        %21456 & 
        \includegraphics[width=.3\linewidth]{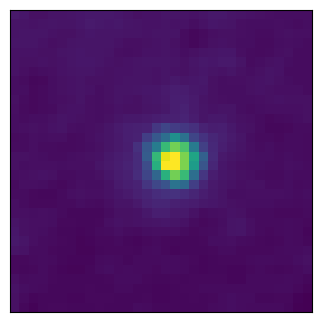} & \includegraphics[width=.3\linewidth]{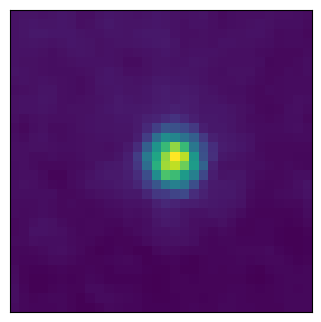} & \includegraphics[width=.3\linewidth]{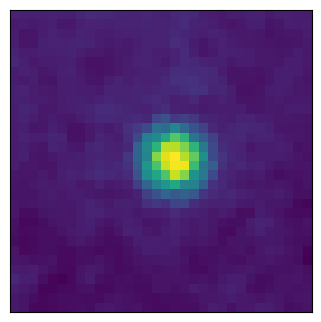} \\
        %21722 & 
        \includegraphics[width=.3\linewidth]{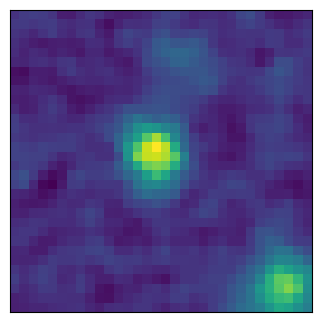} & \includegraphics[width=.3\linewidth]{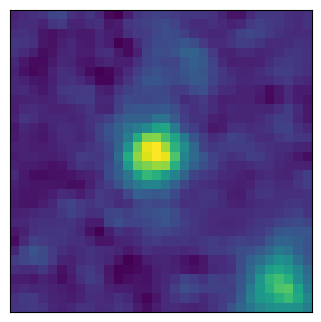} & \includegraphics[width=.3\linewidth]{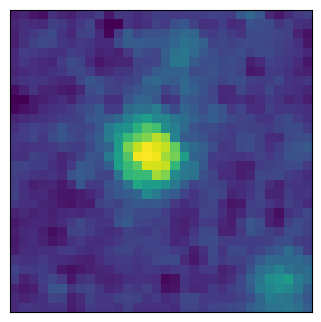} \\
        %31022 & 
        \includegraphics[width=.3\linewidth]{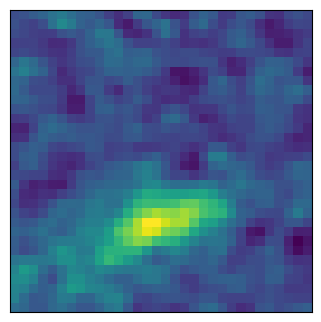} & \includegraphics[width=.3\linewidth]{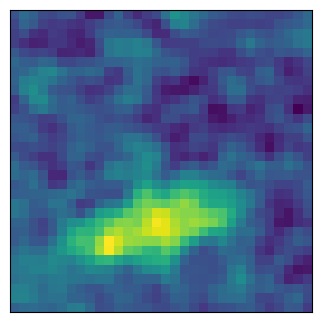} & \includegraphics[width=.3\linewidth]{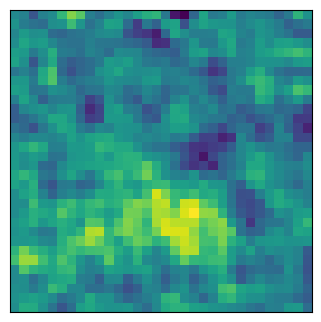} \\
        %2773 & 
        \includegraphics[width=.3\linewidth]{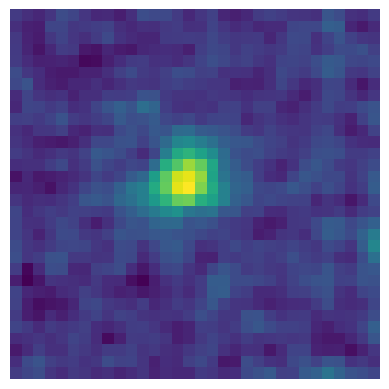} & \includegraphics[width=.3\linewidth]{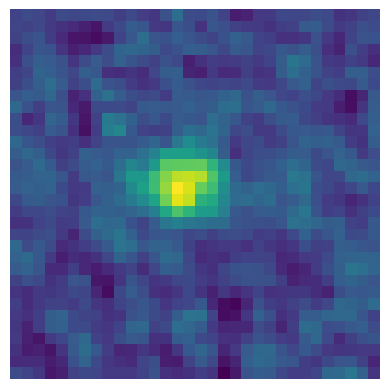} & \includegraphics[width=.3\linewidth]{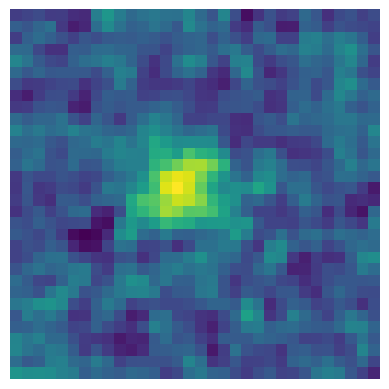} \\
        %8079 & 
        \includegraphics[width=.3\linewidth]{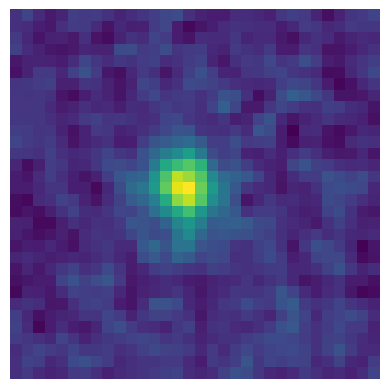} & \includegraphics[width=.3\linewidth]{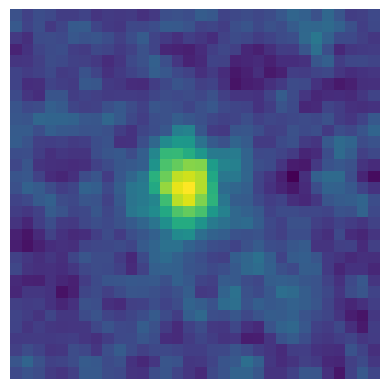} & \includegraphics[width=.3\linewidth]{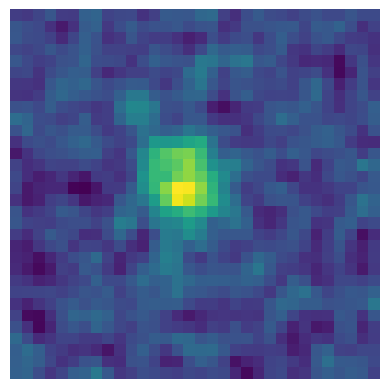} \\
        %14177 & 
        \includegraphics[width=.3\linewidth]{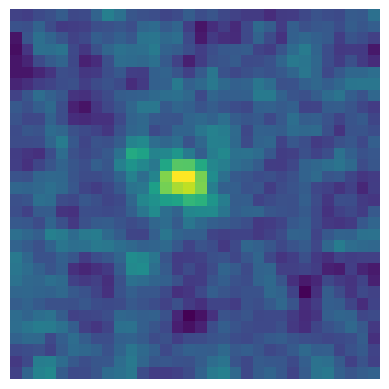} & \includegraphics[width=.3\linewidth]{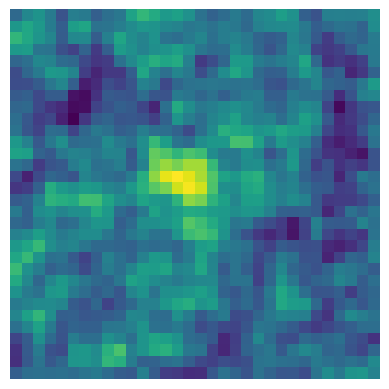} & \includegraphics[width=.3\linewidth]{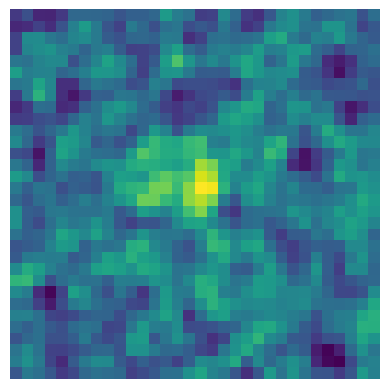} \\
    \end{tabular}
    \caption{\textbf{JWST Images : } A sample of galaxies taken by JWST showing the F277W, F356W and F444W filter images used to test the CNN models. Galaxy samples are taken from the CEERS and JADES data release, specifically of \textit{z} > 6.}
    \label{tab:jwst}
\end{figure}

\subsection{JWST Data} 

We use JWST data to test our results and to determine how well our calibrated machine agrees with different methods for finding dust content values, whether that be the $E(B-V)$ or $A_{V}$ measurements. The imaging and spectroscopic data we use to do this test come from the CEERS survey ID: 1345, PI: S. Finkelstein, see also \citep{Finkelstein2022, Finkelstein2023}. The CEERS field imaging consists of 9 NIRCam pointings with 7 photometric bands (F115W, F150W, F200W, F277W, F356W, F410M and F444W).  We also only use galaxies in this analysis which have confirmed NIRSpec spectroscopic redshifts and spectra within the JADES and CEERS fields, which we have described and analysed in detail in \citet{Duan2023}.

In this paper we utilise our own bespoke reduction of the imaging data from the Cosmic Evolution Early Release Science Survey in the Extended Groth Strip field (EGS).  We have reduced this data independently ourselves using a custom set-up of the \textsc{JWST} pipeline version \textsc{1.6.2} using the on-flight calibration files available through the \textsc{CDRS 0942}. An extensive description of this process and the resulting data quality in provided in \cite{Ferreira2023b, Adams2023}.

Discrepancy between training data and testing data, known as dataset shift, potentially weakens the capability of generalisation of our machine learning model. Testing our machine on JWST data helps determine how well it can generalise to other data, exposing issues in overfitting but also showing if models trained on simulated data can be applied to observed data. The sample of galaxies is taken from systems that are spectroscopically confirmed at $z \sim 7$. We compare our predictions to SED fitted values using \textit{Bagpipes} from \citet{Adams2023} and \citet{Harvey2024}. Below we give a more detailed description of the data sets from which the imaging we use is taken from.  A further description of this data is presented in \citet{Duan2023}.

\subsubsection{CEERS NIRCam Imaging}

Imaging we use from the CEERS (CEERS; ID=1345) NIRCam imaging \citep{bagley2023ceers} includes data across the filters: F115W, F150W, F200W, F277W, F356W, F410M, and F444W. These data have a $5\sigma$ depth of around 28.6 AB magnitudes within 0.1 arcsec circular apertures \citep[][]{Adams2023}.    The data we analyse was collected during June 2022, accounting for 40\% of the total NIRCam area coverage.

We use our own bespoke reduction of this data using a custom set-up of the \textsc{JWST} pipeline version \textsc{1.6.2} using the in-flight photometric zero-point calibration files available through the \textsc{CDRS 0942}.  We provide an extensive description of this process and the resulting data quality in \cite{Adams2022, Adams2023}.

In addition to this, we use the v1.9 EGcS mosaics HST data from the CEERS team, processed using the methodologies outlined in \cite{2011ApJS..197...36K}. This notably includes enhancements in calibration and astrometric accuracy beyond the default HST archival pipeline, within a pixel scale of 0.03".  Within the HST imaging, two filters -- F606W and F814W -- are used in our analyses due to their superior spatial resolution and depth when compared to HST/WFC3 images.  These bands also are bluer than the JWST data and thus provide a blue base line for our SED fitting.  It is clear that using these two HST filters within CEERS is critical for measuring accurate redshifts and other physical properties as this JWST dataset is missing the important F090W band \cite{Adams2022}.

\subsubsection{JADES NIRCam Observations}

The JADES field was observed with NIRCam \citep{rieke2023jades} which cover both the GOODS-S and GOODS-N fields. The galaxies we use in this paper come from the GOODS-S field data (PI: Eisenstein, N. Lützgendorf, ID:1180, 1210). The observations of this field utilise nine filters: F090W, F115W, F150W, F200W, F277W, F335M, F356W, F410M, and F444W, over an area of about  25 arcmin\(^2\).   Within these observations a minimum of six dither points were used with exposure times ranging from 14-60 ks. We calculate that the $5\sigma$ depths range from 3.4 to 5.9 nJy, within aperture sizes between 1.26 and 1.52 arcsec. Across all filter bands, JADES ensures a high level of pixel diversity \citep{rieke2023jades}, thereby significantly reducing the impact of flat-field inaccuracies, cosmic ray interference, and other issues at the pixel level. In this paper, we utilize the publicly released JADES data and reductions.

\subsubsection{SED Fitting}

We use the JWST imaging data not only for the morphology and structure for the Machine Learning applications, but also the SEDs constructed from the JWST+HST bands. We fit these SEDs using the {\it Bagpipes} \citep{Carnall2018, Carnall2019} code to both the photometric and spectroscopic data for our sample of systems.  We do this for six parametric SFH models, including: log-normal, delayed, constant, exponential, double delayed, and delayed burst—along with a non-parametric Continuity model \citep{leja2019measure}. These models are widely accepted and have been used in various works \citep[e.g.][]{2023Natur.619..716C,2023MNRAS.524.2312E,2023MNRAS.522.6236T, 2023MNRAS.519.5859W, 2023arXiv230602470L}. 

We carry out these fits by fixing the redshift to the spectroscopic redshift in all cases throughout this paper.   Within these fits, we use Log10 priors for the dust properties we fit.   We select Log10 priors as we expect high redshifts galaxies to be young, with a lower metallicity. We set prior limits for metallicity in the range of $[1\text{e-06}, 10.0] \, \text{Z}_{\odot}$, dust prior in the range of $[0.0001, 10.0]$ in $A_\text{V}$, ionization parameter $\mathrm{Log}_{10}(\mathrm{U})$ in range of $[-4, -2]$, the star formation start time is placed at 0.001 Gyr, the time assumed for star formation to stop is at $t_\text{U}$ -- the age of the Universe at the observation. We use a Kroupa IMF as well as \cite{2003MNRAS.344.1000B} SPS models, and the  \cite{Calzetti1994} dust attenuation model is implemented. As we use dust attenuation inferred from SED fitting, which is not significantly affected by different SFHs, we have chosen to present results inferred from our fiducial model, the log-normal SFH.  The one limitation to using these SED fits is that we assume the dust extinction law is Calzetti.  We largely keep this as this is by far the most common dust extinction line used to measure the photometrically derived properties of galaxies and one goal of this paper is to determine how accurate this assumption is based on an independent derivation.

\subsection{Deep Learning CNN Model}

\subsubsection{Data Augmentation}

Whilst we have many simulated galaxies from the Illustris TNG, the more imaging used in the training the better in terms of performance for a CNN model.  We thus carry out an augmentation of modifying our existing origianl sample to create a large one derived from this original imaging. 

To increase the number of images for our use, the simulated TNG images were augmented to increase the training set size. First, each image was rotated by 90, 180 and 270 degrees clockwise. The training dataset was then balanced to have a more even distribution of predictive values. Augmentation of images included : image rotation, image flip across the x or y-axis, addition of Gaussian noise, or applying a histogram of gradients (HOG) to the image. Without augmentation, outlier values are often ignored, as it is a small subset of the data \citep{Longadge2013}. We also want to use augmentation to balance the distribution of the values in the dataset. As the dataset has a Gaussian distribution, the machine is more likely to estimate values close to the median of this dataset as this will have a small loss value, thus appearing to be more accurate. Using a loss function such as mean squared error (expanded upon in later sections) penalizes the machine more for wrong guesses at the extreme, and so over the course of training the machine becomes more conservative in its predictions. To account for this, we raise the proportion of large values in the training set.

The augmentation of the smaller datasets helps supplement the amount of images available for training and can hopefully lead the machine to be able to predict and generalize to a wider range and variety of images. We want to use good quality images to ensure better training results, so we impose a signal-to-noise ratio (SNR) cut-off of SNR $\geq$ 5 on the images of galaxies to get a set of clean data. Imposing this SNR cut-off cuts down the original images we have to use from 33,376 to 6,620. Augmentation helps increase the amount of usable images to 25,208. The set of images is split into two groups, 85$\%$ in the training set and 15$\%$ in the testing set, where the training set is defined as the dataset used to train the machines. During training, a smaller subset of the training set of data is further split from the main training data. After each epoch of training, the CNN model will test the machine against this smaller subset of data, known as the validation set \citep{Wang2016}. A validation set is used as a proxy for the testing set inside the training period. As the model has not seen the validation set of images in the training epoch, it can be used to estimate whether or not the machine should be early-stopped. On the other hand, the testing set is used after training to assess how well the machine performs as it is a set of images that the machine has not seen at all during training.

\subsubsection{CNN Model Set-up}

The Convolutional Neutral Network (CNN) model built for this analysis uses the tensorflow and keras package. To optimize the tuning algorithm, we used the keras-tuner package. The CNN model consists of two main sections, the convolution layers and the neural network layers. A convolutional neural network is named after the convolution layers, which use a number of different matrices to select features of the images, such as borders of objects. This layer uses the convolution operation to convolve two sets of data, here being the input and the filter, to create a feature map. This feeds into the fully-connected layers which works on sorting inputs or calculating properties (For more details on CNNs see \citet{Albawi2017}). 

There are two main variables in a convolution block, the number of filters in each layer and the number of layers. We use a fixed kernel size of 3 by 3 units for the convolution layers, followed by a \texttt{BatchNormalization} layer and a 2 by 2 \texttt{MaxPooling} layer. The kernel, or filter, mentioned is used to convolve the input data. Filter weights are finetuned in the training process so the convolution process shows features from our images. The \texttt{Pooling} layer is another process to help particular features stand out by aggregating pixels together to maximise features. Previous work from \citet{Cheng2020} used \texttt{AveragePooling}, but here we find that the loss is smaller when \texttt{MaxPooling} was used. The \texttt{BatchNormalization} layer works to normalize the input distribution during training to prevent large internal shifts, known as internal covariate shift \citep{Ioffe2015}.   An activation function is applied as a non-linear operation after the CNN layers and we use the rectified Linear Unit (\texttt{reLU}), which follows the formula $f(x) = max(0,x)$, a common activation function that has shown to perform well for CNNs \citep{Agarap2018}.

The data is then flattened into a 1D array before being passed into the neural network layers, referred to as \texttt{Dense} layers below. Neural networks consist of different number of layers that assign weights to features found by the convolutional layers in order to predict an expected value. The number of filters in the layers determines the complexity of the model. Right before the output layer, we employ a dropout layer which shuts off a percentage of the output nodes (50$\%$ here). This is done to prevent outputs from coming from only a small subset of nodes and spreads the learning amongst all nodes, a common approach to prevent overfitting \citep{Srivastava2014}. This dropout layer is used in training only, so all output nodes are used during evaluation of testing data, which should help performance when the model predicts using testing data. 

We utilise hyperparameter training to help determine the optimal value for each variable in our model. The complexity of neural network machines mean that there are a large number of variables and an even larger variable space to explore. Each convolution and dense layer includes variables such as the number of filters, activation function, batch size, and even kernel size. Within the structure of the machine there are even more variables such as the choice of learning rate, learning algorithm, loss values, and activation functions. In hyperparameter training, variables are given a range and step, which will be fine-tuned over the hyperparameter training process to give the smallest loss value \citep{Yu2020}. For example, the number of filters in a convolution layer was set within a range between 4 and 64. The hyperparameter tuning runs through different combinations of parameters to try to pinpoint an optimal set of parameters that could perform the best on the testing data. 

Other hyperparameters included the learning rate of the optimizer function {\it Adam}, the Adaptive Movement Estimation algorithm (Introduced in \citet{Kingma2014}). Values were allowed to vary between 0.01, 0.005, 0.001, 5~$\times 10^{-4}$ and 1~$\times 10^{-4}$, but centered on 0.001 as recommended as per \citet{Kingma2014}. The learning rate helps determine how much the machine tunes its weights in response to results of each trial. Large learning rates can cause more variability, leading to longer training times and reach sub-optimal results. However, small learning rates requires a longer time for the machine to move through the variable space \citep{Wilson2017}. In tuning, the different learning rates are tested and the learning rate which results in lower evaluation loss values is used. The tuned hyperparameters we use are listed in Table \ref{tab:param}. Performance is assessed by the loss function. Here we use the percentage error function. Other common loss functions, such as mean squared error, yield small error values due to the small values and cause poor training results. Keras-tuner offers multiple optimization methods, where we chose to use the Bayesian Optimization algorithm. The other optimization methods, such as grid-search or random-forest walk, took a longer time to traverse the variable space in comparison.

\begin{figure}
    \centering
    \includegraphics[width=\linewidth]{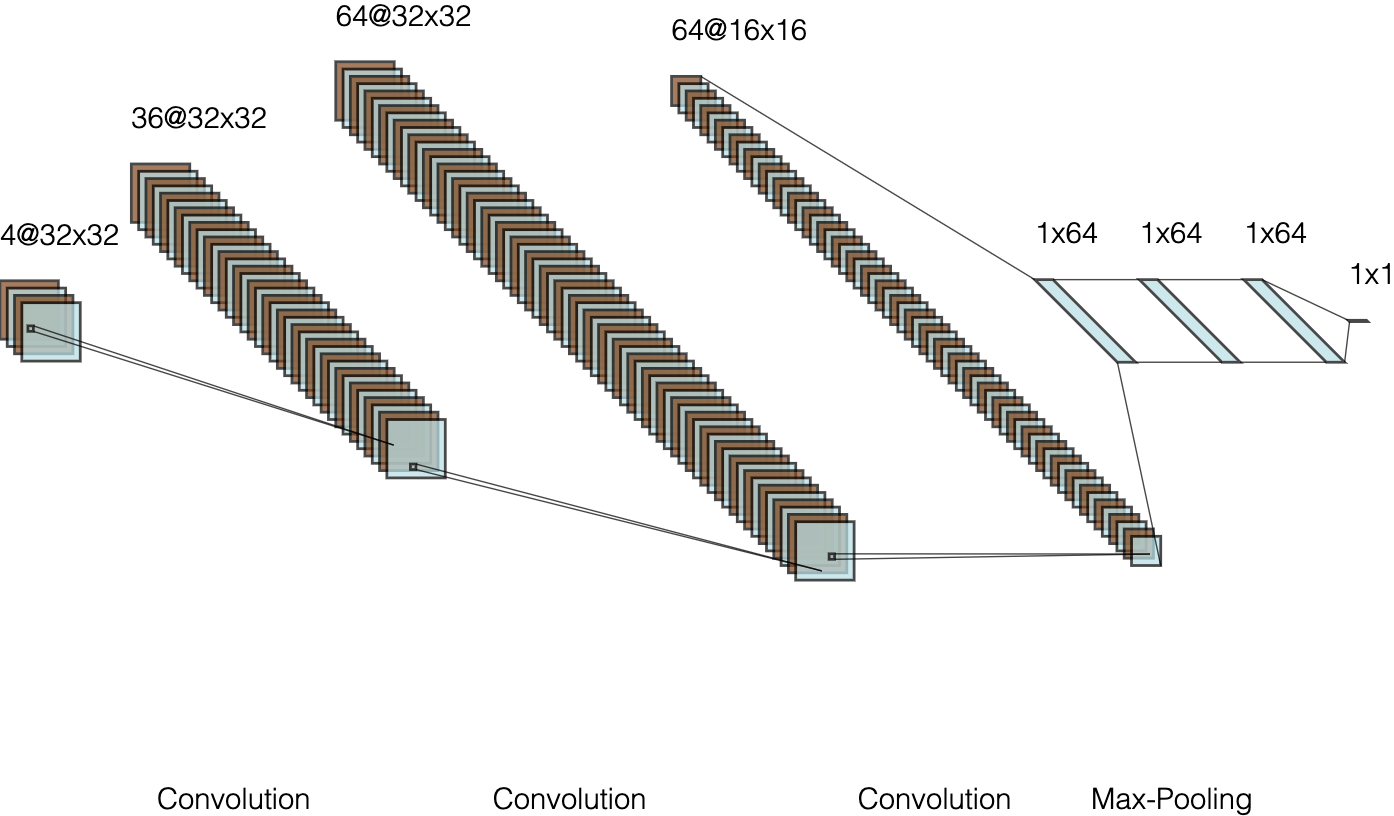}
    \caption[Machine Structure]{\textbf{Machine Structure}: Machine structure for the one band input for the machine predicting E(B-V) values. The numbers written on top of the CNN layer gives the dimensions of the particular layer. For example, 36@32x32 indicates 36 layers of dimensions 32 by 32. For the Dense layers on the right hand side, 1x64 means the layer has 64 nodes.}
    \label{fig:1ebvmachine}
\end{figure}

\begin{table}
    \centering
    \begin{tabular}{l | c | c}
       Parameter & Range & No. of layers \\
       \hline
       CNN filters & 4 - 64 & 1 - 4 \\
       Dense neurons & 8 - 64 & 1 - 4\\
       Learning Rate & 0.01 - 0.0001 & - \\
       Batch Size & 16 - 64 & - \\  
    \end{tabular}
    \caption{\textbf{Hyperparameters}: The range of parameters used in hyperparameter training, tuned to optimize accuracy. All CNN and Dense layers in a machine structure are given the same range. During hyperparameter tuning, the optimization algorithm tests different sets of parameters to find the set that returns the lowest loss value.}
    \label{tab:param}
\end{table}

Observations at different wavelengths can show different properties. We use three bandpass images (F277W, F356W, F444W) to utilise these distinct morphological features. In a three-input model, each image is taken into an individual convolutional block and then condensing the three inputs into one array for the dense layers. We expect better predictions as the wavelength dependency of dust extinction should show up and be compared across different bandpass images. A drawback is that optimizing for multiple inputs increases the number of parameters to tune and requires more time and memory needed during training. We set 200 trials for the hyperparameter tuning process, where each trial tested a different set of parameters on the machine. Training is cut short if the validation loss did not improve over 5 epochs of training. This helps cut down on the time needed as poorly performing models are terminated early. We chose the machine structure of three CNN layers and four dense layers which resulted in the lowest loss values for $E(B-V)$, illustrated in Figure \ref{fig:1ebvmachine}. Due to time constraints, we used the same configuration for models for the other properties. 

\begin{figure*}
    \centering
    \includegraphics[width=\linewidth]{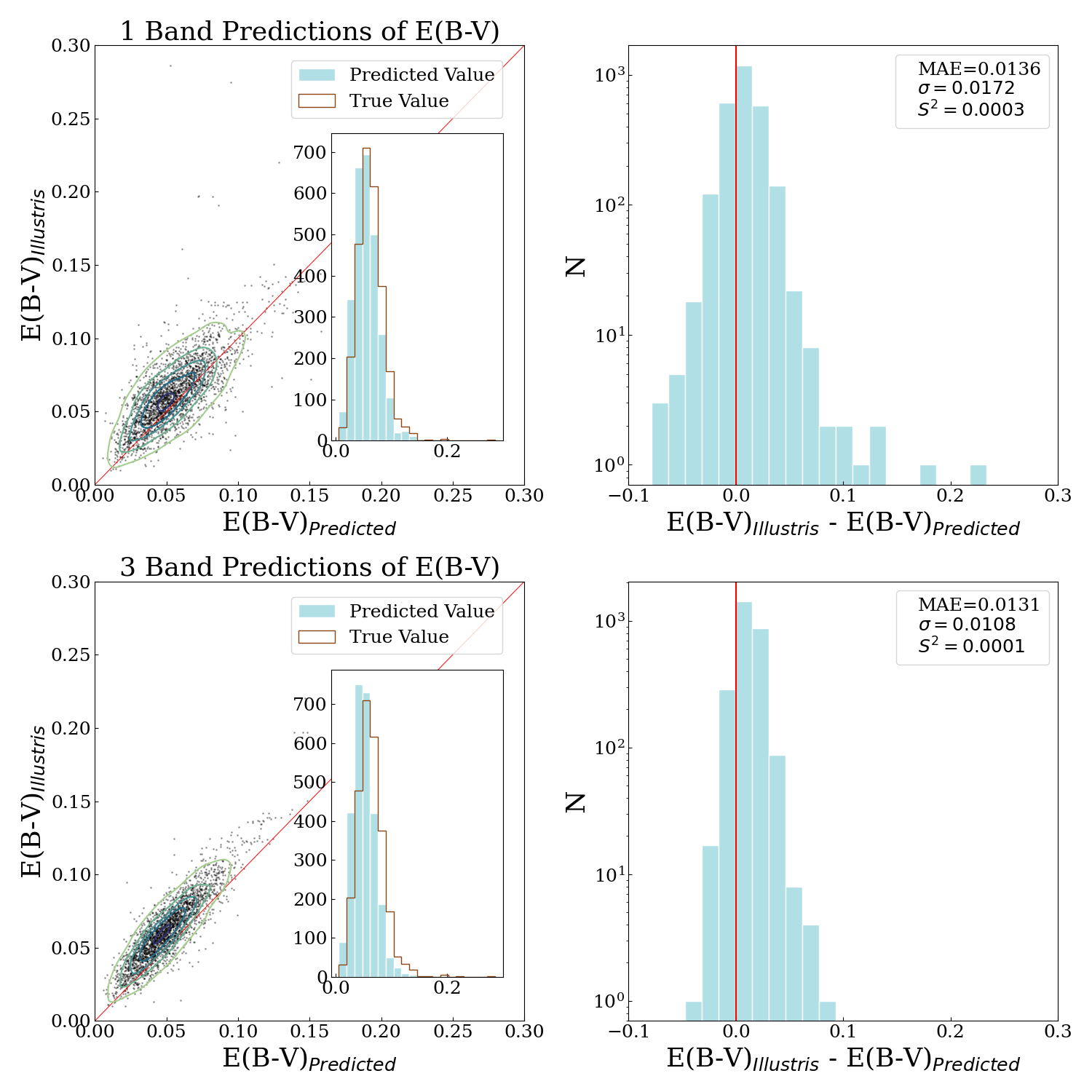}
    \caption{\textbf{Machine learning E(B-V) Predictions :} \textbf{(Top)} Comparison of the simulated 'true' $E(B-V)$ values against the machine predicted results with one band input on the top, and the three band inputs on the bottom. The red line charts where the predicted value is equal to the true value, and deviation from this line shows a bias in the distribution. We show the distribution of the predicted value in blue bars and the true value in a red outline in the histogram. On the right hand side, the histogram of the difference between the true and predicted values, calculated by subtracting the predicted from the true value. A longer trailing tail towards the right of the distribution show a bias of the machine towards under predicting values, which can also be seen in the difference in overall distribution on the small histogram and scatterplot on the left. 
    \textbf{(Bottom)} The scatterplot is more concentrated with less scatter when compared to that of the one input, but still shows under-prediction at larger values when the scatter deviates away from the red line. This also manifests in the histogram of differences on the right, which is skewed slightly right of the red $x=0$ line. Compared to the one input histogram on top, the differences from the three input are generally smaller, indicating a better fit.}
    \label{fig:ebv}
\end{figure*}

\begin{figure*}
    \centering
    \includegraphics[width=\linewidth]{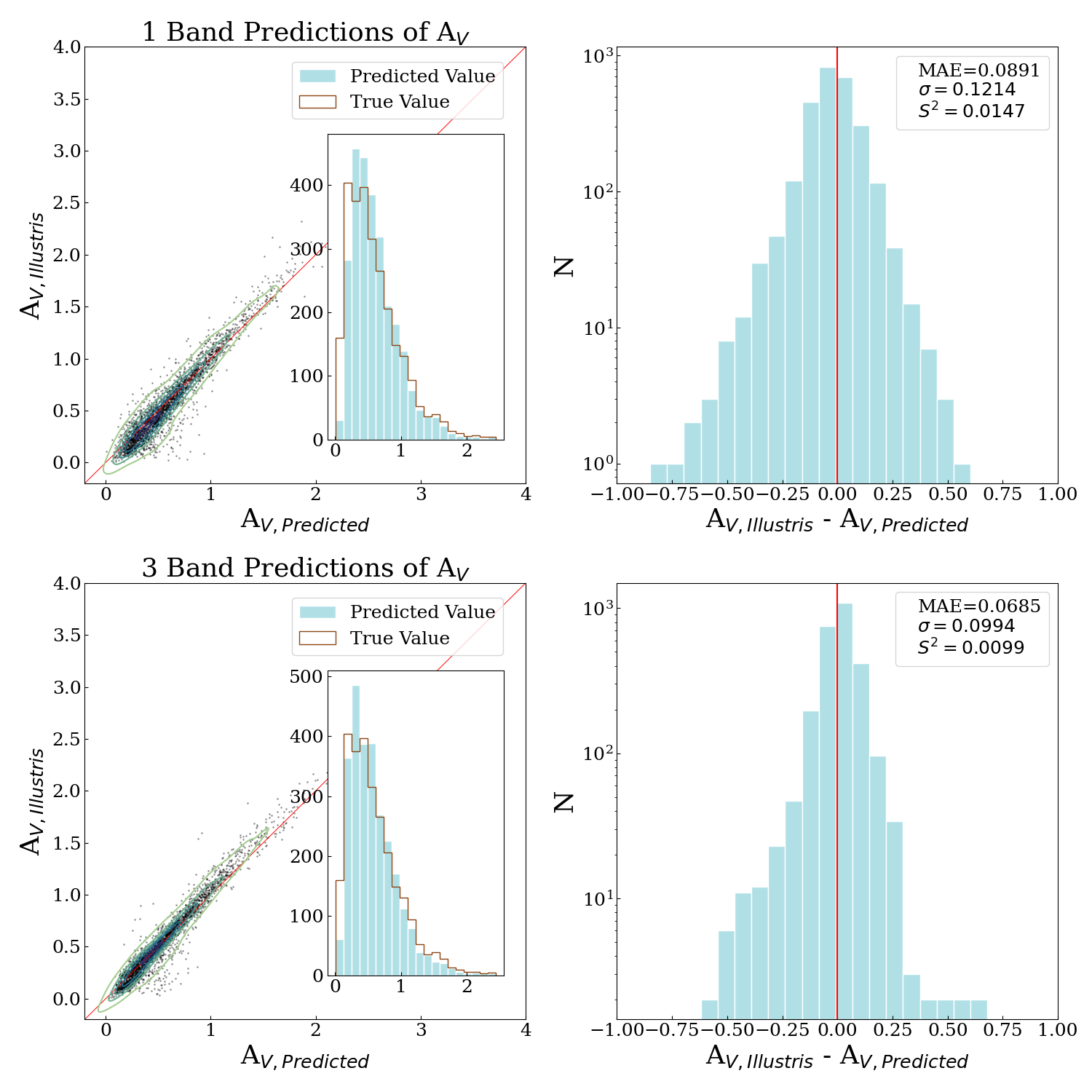}
    \caption{\textbf{Machine learning $\textup{A}_{V}$ Results :} \textbf{(Top)} Comparison of the machine predicted $A_{V}$ results against the simulation based 'true' values of dust extinction $A_{V}$, with the two plots at the top showing results from the one band model. On the left scatter plot, predictions cluster around the red (true = predicted) line but our ML model underestimates some values at the extremes. The distribution of true values is skewed more towards the left than the distribution of the predicted values. This is reflected in the distribution of their differences on the right, where the histogram has a very slight tail towards the left, showing that the predicted values tended to be larger on average than the expected true value.  Therefore our ML method is slightly over predicting the correct $A_{V}$ values for galaxies based on its training. 
    \textbf{(Bottom)} Predictions from the three band model. The left scatterplot shows a tighter distribution around the red line as compared to the one band model. A smaller histogram comparing the distribution of predicted and true values show similarly shaped distributions. On the right, the histogram of the differences shows that the errors are slightly better than the one input model on the top. However, there is also a tail towards the left, showing a similar tendency of over-prediction of dust extinction as the one input model.  As can be seen the three band model has a tighter distribution of differences from true minus prediction, but the distribution is more skewed. }
    \label{fig:av}
\end{figure*}

\begin{figure*}
    \centering
    \includegraphics[width=\linewidth]{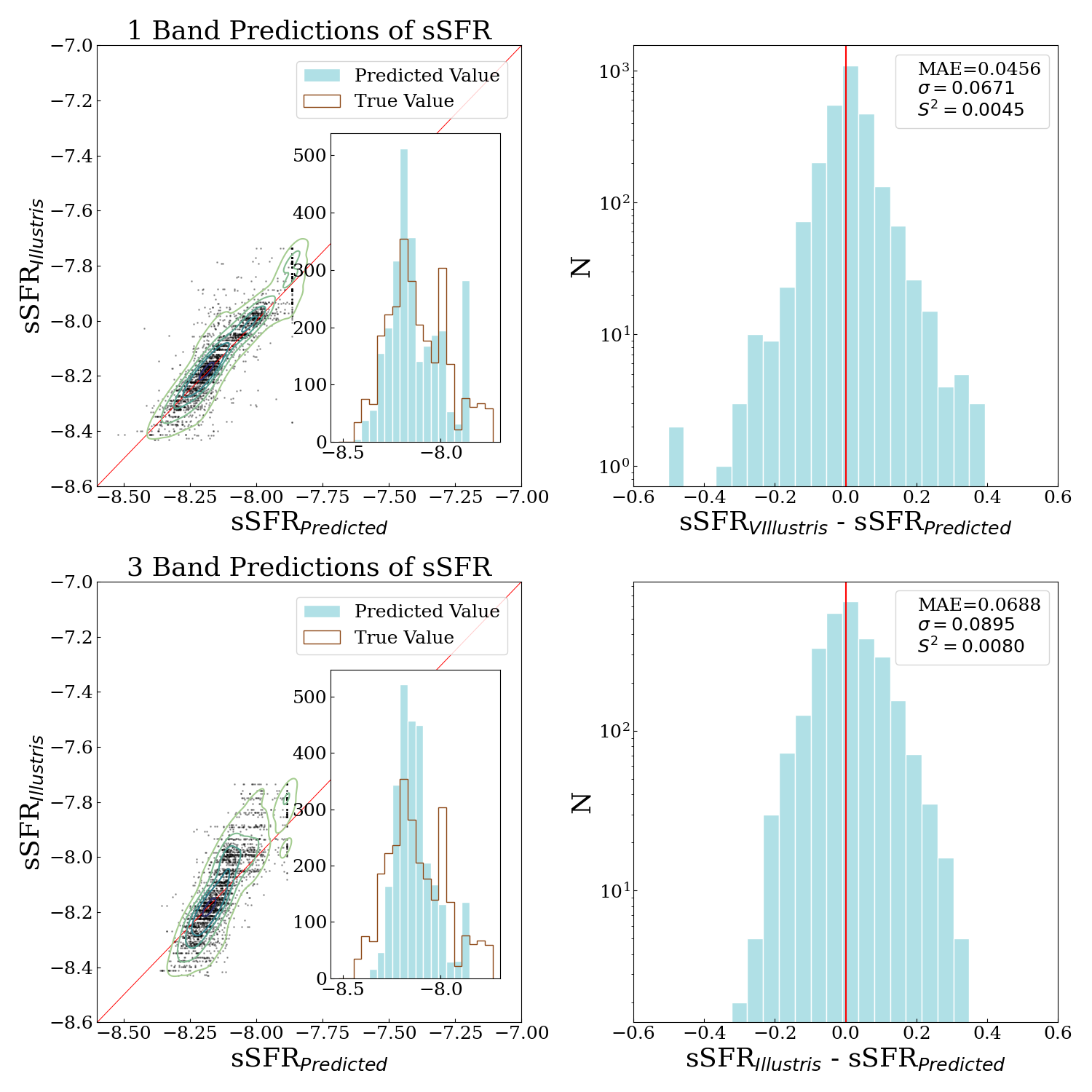}
    \caption{\textbf{sSFR Results : (Top)} Comparison of the machine predicted results against the true values of sSFR in the units of log (yr$^{-1}$). The distribution of predicted values for the one input has a limited range and is concentrated towards the median. There is clearly a sharp line at the largest value that the model can predict, seen in the scatterplot and histogram of predicted values. On the right, there are a number of outliers, but the overall distribution of the differences between true and predicted values is more symmetrical. 
    \textbf{(Bottom) } The scatter strength for the three input model for sSFR is clearly weaker, and the distribution shows that the model predictions have a smaller range than the one input model and the model over-predicts at the median values. There is also a similar line where the machine is not able to predict any larger values. The difference between true and predicted values is similarly symmetrical. However, the average difference is larger than that of the one band input.}
    \label{fig:ssfr}
\end{figure*}

\begin{table*}
    \begin{tabular}{l | cc | cc | cc | cc}
        \hline
        \multicolumn{8}{c}{Evaluation of Predicted Values} \\
        \hline
         & \multicolumn{2}{c}{E$_{(B-V)}$} & \multicolumn{2}{c}{A$_{V}$} & \multicolumn{2}{c}{R$_{V}$} & \multicolumn{2}{c}{sSFR} \\
        \hline
         & 1 Band & 3 Band & 1 Band & 3 Band & 1 Band & 3 Band & 1 Band & 3 Band \\
        MAE & 1.36 $\times 10^{-2}$ & 1.31 $\times 10^{-2}$ & 8.91 $\times 10^{-2}$ & 6.85 $\times 10^{-2}$ & 3.4817 & 2.8892 & 4.56 $\times 10^{-2}$ & 6.88 $\times 10^{-2}$ \\
        MSE & 3.55 $\times 10^{-4}$ & 2.53 $\times 10^{-4}$ & 1.52 $\times 10^{-2}$ & 9.94 $\times 10^{-3}$ & 45.9 & 18.7 & 4.60 $\times 10^{-3}$ & 8.16 $\times 10^{-3}$ \\
        Standard Deviation $\sigma$ & 1.72 $\times 10^{-2}$ & 1.08 $\times 10^{-2}$ & 0.121 & 9.94 $\times 10^{-2}$ & 6.2325 & 3.4815 & 6.71 $\times 10^{-2}$ & 8.95 $\times 10^{-2}$ \\
        Variance $S^{2}$ & 2.95 $\times 10^{4}$ & 1.17 $\times 10^{-4}$ & 1.47 $\times 10^{-2}$ & 9.89 $\times 10^{-3}$ & 38.8441 & 12.1205 & 4.50 $\times 10^{-3}$ & 8.01 $\times 10^{-3}$ \\
    \end{tabular}
    \caption{\textbf{Comparison of Predicted Values : } To compare and evaluate the difference between the 1 input and 3 input machines, we look at how well the machine performed overall, illustrated in the differences between the true and predicted values. We compute the mean average error, mean squared error, standard deviation, and variance of the distribution of the differences to compare how well the one-input machine performs to the three-input for each different value. Overall, the three-input model seems to perform slightly better than the one-input model.}
    \label{tab:val}
\end{table*}

\section{Results}
\label{sec:results}

\subsection{Training}

We train and test our CNN models on three different quantities of interest ($E(B-V)$, $A_{V}$, sSFR) using one (F277W) and three (F277W, F356W, F444W) images as input, for a total of six different tests run. A set of values of $R_{V}$ is calculated using the relationship from Equation \ref{eq:rv}. Our aim is to determine how well the visual image and the morphological features of a galaxy can be used to calculate its dust properties. By using a single image and three images, we can also investigate the potential of a CNN model in evaluating morphological features between different wavelengths for measuring the dust content. 

The result from training the machine with one input (F277W) is shown in Figure \ref{fig:ebv}. We compare the true and predicted values on the left, plotting the values against each other in a scatterplot and comparing the distribution in the histogram on the right. The shape of the distribution indicates that the trained machine tends to under-predict actual values, and the scatterplot of values show a good linear correlation between both values upon the identity function ($x=y$) represented in a red line. A set of good predictions should follow this red line, whereas our one input machine results show a wide clustering around this red line and a couple of outliers. Overlaying the scatterplot is a 2D probability density plot drawn to include [10, 30, 50, 70, 95] percent of the total values which helps illustrate how clustered these predictions are. 

For the one input $E(B-V)$ model, the scatterplot and probability density contours show a slighlty wide spread. The density contours show that the majority of the points lay towards the left of the red line, showing that our model generally under-predicts values. This is shown in the small histogram in the same plot, where the peak of the blue histogram (Predicted Value) is to the left of the brown histogram (True Value). To evaluate how well our machine can predict colour excess, we look at the difference between the true and predicted value on the histogram on the right. This distribution goes to zero if the predicted and true values are the same, represented by the red line. 

To quantify these differences better we calculate the mean absolute error (MAE) to illustrate how different the predictions and the expected value are using

\begin{equation}
    MAE = \frac{1}{N} \sum^{N}_{i=1} |y_{i} - \hat{y}_{i}|
\end{equation}

\noindent The value of MAE is unable to differentiate if the model under-predicts or over-predicts by itself as it has no direction. Here, the distribution of the differences skews left, showing a non-zero (0.0136) MAE for the differences and indicating that the machine tends to under predict the colour excess. Overall, the distribution is relatively symmetric aside from the tail skewing left. Predictions from the three input machine shows similar values, such as a wider clustering and under-prediction. From Table \ref{tab:val} the average differences for the 3 Band predictions of colour excess is $= 0.013$, which is very close to the results from the single band trained machine. Conversely, the distribution from the three Band predictions have a smaller standard deviation and variance, ($\sigma = $ 0.0172 to 0.0108). We observe similar trends in other values of the dust extinction and sSFR. 

Figure \ref{fig:av} compares the results of dust extinction values ($A(V)$) using one input and three input. The predictions of $A_{V}$ show tighter clustering as compared to values of $E(B-V)$ from the scatter plot and density contours with less numbers of outliers. The histogram shows a slight right-leaning bias, indicating that there is also an over-prediction. This is also evident in looking at the histogram of the differences between the true and predicted values which has a trailing tail to the right. Results from using three bands to predict dust extinction shows tighter clustering and more accurate predictions as compared to the one band results. The distribution of the differences of one and three input predicted values has a smaller average (MAE = 0.09 to 0.07) and a smaller standard deviation (MAE = 0.012 to 0.099) compared to the differences using one input. Overall, the range of differences is slightly larger for the one input than the three input. 

With both values of $E(B-V)$ and $A_{V}$, we can calculate the R-value using Eq. \ref{eq:rv}. As the $R_{V}$ value is calculated using two predicted values, we expect larger errors. This is reflected in the larger spread of $R_{V}$ values. The shape of the distribution shows an over-prediction, also seen in the histogram of differences, skewing left. This may be due to the general trend of smaller than predicted $E(B-V)$ values causing large $R$-values. While the three input model is comparatively better, the same trend is observed. Differences between the true and simulated $R_{V}$ values from the three input model are overall smaller than that of the one input, as well as a smaller average (MAE = 3.48 and 2.89) and standard deviation ($\sigma = $ 6.23 to 3.48), possibly due to the one input models for both $E(B-V)$ and $R_{V}$ have larger errors on average compared to the three input models. 

The final value we test and measure is the specific star formation rate (sSFR), represented in log$_{10}$  yr$^{-1}$ units. Figure \ref{fig:ssfr} shows the results of the machine training.  
 One thing to immediately take away form this is that the ML predictions for the sSFR are not as good as for the dust extinction values.  Star formation remains a challenging feature to measure with machine learning methods as also found in other papers such as \citet{Bisigello2023}.  Another odd feature of the predictions is the inability to predict larger sSFR values as there is a sharp line on both one and three input models in both the scatterplot and the histogram. While there are not many outliers in the predictions from the scatterplot, the correlation is not as strong as compared to those for dust extinction or colour excess. The distribution of the differences between the true and predicted values is quite symmetric without a tail in either direction. It is interesting to note that the sSFR is the only property tested where the three input model did not perform better than the one input. Both the average difference and standard deviation from the one input are smaller than that of the three input, indicating that even though the range of differences is larger from the one input, the distribution is more concentrated.  This type of regression analysis has been shown to work well on imaging, but also using photometry, to measure other galaxy properties, such as structural properties, redshift, stellar mass ,and star formation rates \citep[e.g.,][]{Tohill2020, Bisigello2023}

\begin{figure*}
    \centering
    \includegraphics[width=\linewidth]{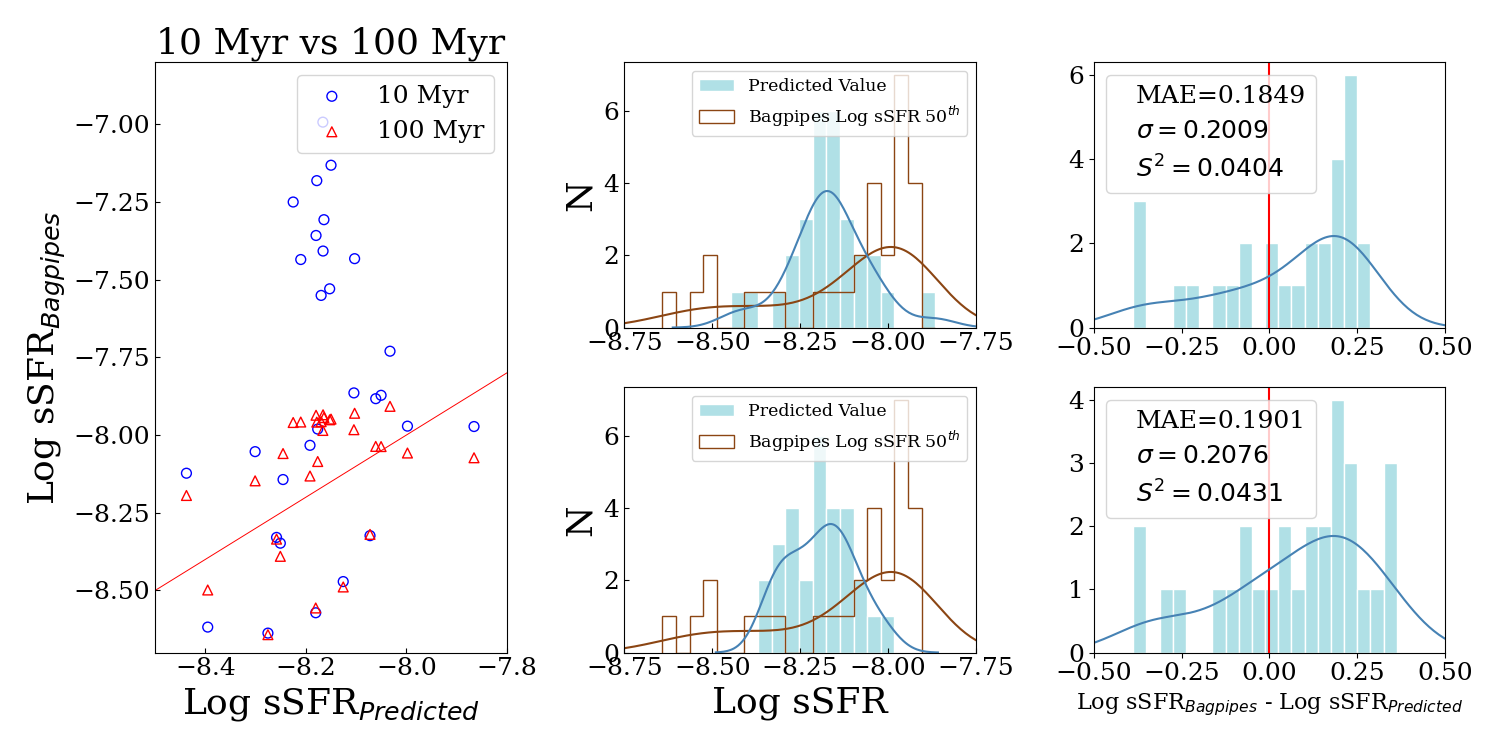}
    \caption{\textbf{Predictions of the specific star formation rate (sSFR) values for the JWST galaxy samples : (Left) } Scatterplot comparing the two sets of values of sSFR generated using the 10 Myr and 100 Myr of star formation in the {\it Bagpipes} code. For sSFR, there is a much larger difference between the two sets of values. Since the simulated images used for training were generated with Illustris using a 100 Myr time-scale, we compared our predictions to that of the 100 Myr set of {\it Bagpipes} data. 
    \textbf{(Middle) } Predictions of sSFR values using the one band model (Top) and three band model (Bottom). Both models can only predict a limited range of sSFR values compared with the {\it Bagpipes} results. It appears that predictions from the one input model are more symmetrical and have a larger peak at the median, compared to the three input model. 
    \textbf{(Right) } The differences between the predicted and {\it Bagpipes} values are shown, with the difference with one input on top and three inputs on the bottom. Both have a similar distribution which skews slightly right, indicating over-prediction by the machine. }
    \label{fig:jwstssfr}
\end{figure*}

\subsection{Application to JWST imaging}

\begin{figure*}
    \centering
    \includegraphics[width=\linewidth]{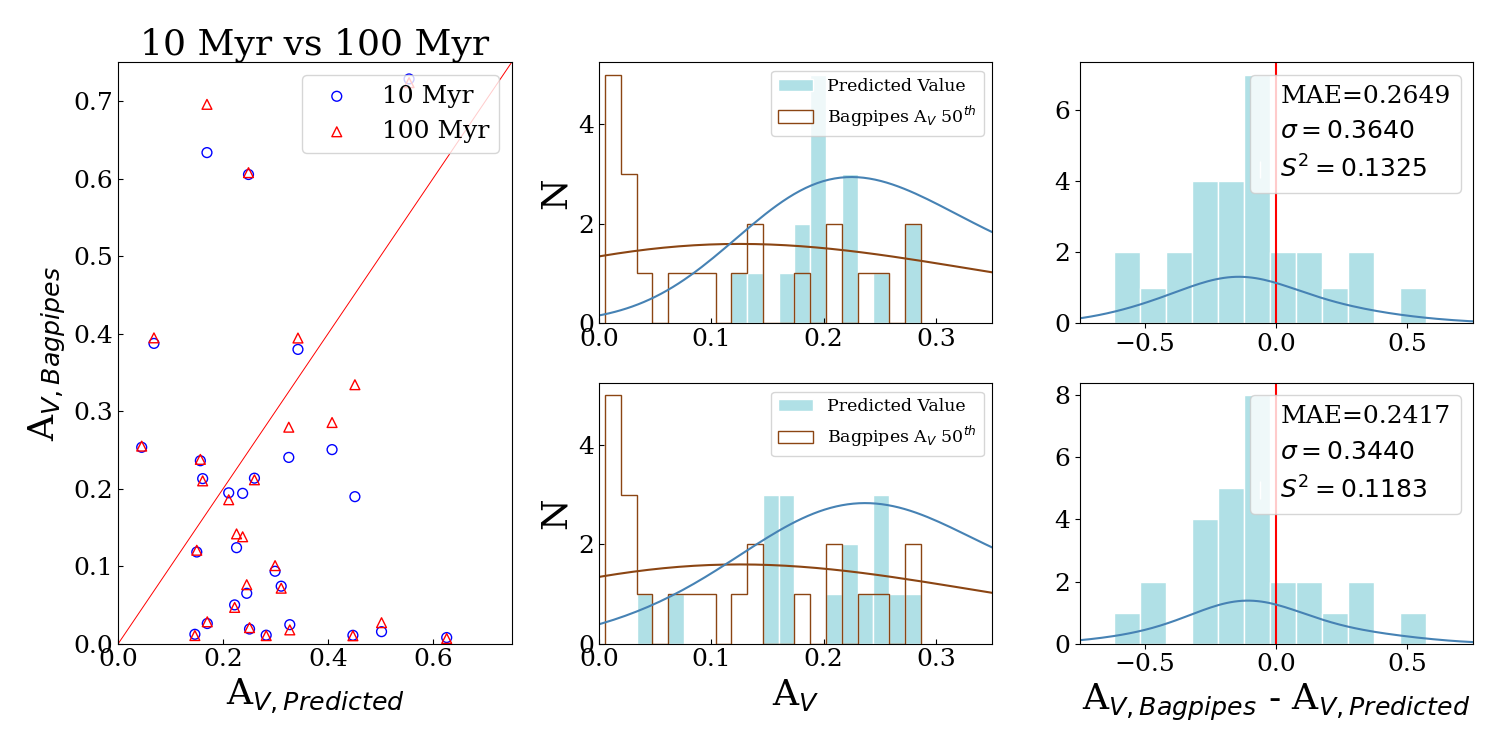}
    \caption{\textbf{Predictions of $A_{V}$ values for the JWST galaxy samples : (Left) } Two sets of values for the physical properties of our JWST galaxy sample as generated using the code {\it Bagpipes}, one assuming a 10 Myr star formation star formation scale and one using a 100 Myr time range. This scatterplot shown here compares the difference between these two sets of values against the predictions by our machine learning method. For dust extinction, there does not seem to be much difference between the two sets of {\it Bagpipes} values, and neither show a strong correlation with our machine predictions. 
    \textbf{(Middle) } The predictions of dust extinction values using the one band model (Top) and three band model (Bottom) on our JWST sample. The blue bars of the predicted values show the machine is predicting large values of $A_{V}$ compared to the SED fitting method, especially at small values. To show the general trend of predictions, a probability density curve (PDF) is overlaid on the histogram, showing that both one and three band models are extremely similar. 
    \textbf{(Right) } The differences between the predicted and {\it Bagpipes} values for the dust extinction, with the difference of one input band shown on top and three input images on the bottom. Both show similar distributions, with a smaller MAE for three input (0.26 vs. 0.24). While the standard deviation and variance for the three input also indicate closer predictions to the {\it Bagpipes} values, differences are slight compared to the one-band predictions. These differences are on the same level as when comparing the {\it Bagpipes} to the Meurer method as show in Fig \ref{fig:bagpipes}.}
    \label{fig:jwstav}
\end{figure*}

\begin{figure}
    \centering
    \includegraphics[width=\linewidth]{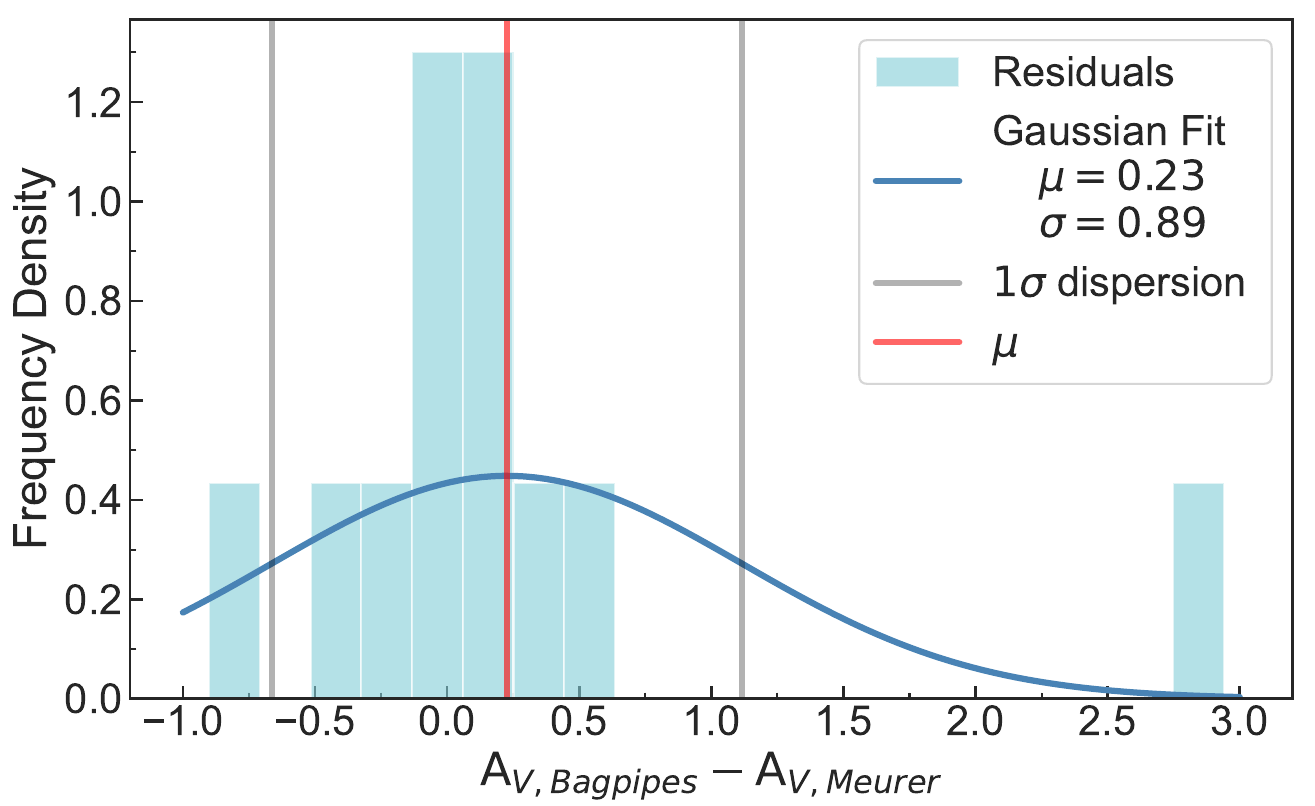}
    \caption[Bagpipes Meurer]{\textbf{Comparison of the \textit{BagPipes} and Meurer relation: } Plot showing the difference in the measurement of A$_{V}$ with the code {\it BagPipes} and those calculated using the Meruer relation. The range of differences between the values of \textit{BagPipes} and the Meurer relation appears similar than the range of values obtained using predictions with the CNN models and \textit{BagPipes}, shown in Fig \ref{fig:jwstav}.}
    \label{fig:bagpipes}
\end{figure}

The release of JWST imaging allows us to test our machine which has been trained on simulated JWST data on real JWST data. To do this, galaxies with a confirmed NIRSpec redshift at $\geq$ 6-7 were selected from \citet{Duan2023}. These are amongst the earliest galaxies which have confirmed spectroscopic redshifts at the redshifts of interest.   Figure \ref{tab:jwst} shows the sample of these JWST images in the CEERS and JADES fields we use. Parameters of dust extinction and sSFR were calculated using the {\it Bagpipes} code using the full SEDs to compare against the predicted outputs by our machine. 
 
Two sets of properties from {\it Bagpipes} are calculated using 10 and 100 Myr star formation time-scales respectively. These time-scales affect the retrieved sSFR values, as can be seen in Figure \ref{fig:jwstssfr}, a scatterplot showing the difference of the two sets of parameters. In the 100 Myr set (red points) there are no values of sSFR larger than -7.75, whereas the 10 Myr SFR set (blue points) has a larger range of sSFR values, up to -7. Since the models were trained on Illustris images generated with a star formation timescale of 100 Myr, we compared our predictions to the set of data using a 100 Myr timescale. 

Figure \ref{fig:jwstav} shows the comparison of dust extinction values when using one band and three band input machine predictions and values from {\it Bagpipes} in the JADES field. On the very left, we compare the set of {\it Bagpipes} data at the 10 and 100 Myr timescales. While the values from using a 100 Myr timescale are, in general, slightly larger than those from a 10 Myr timescale, the differences appear minimal. In the middle, we show the predicted values compared to the {\it Bagpipes} values, with the one band predictions on top and the three band predictions on the bottom. Our predictions are shown in the blue histogram while \textit{Bagpipes} values are shown in a brown outlined histogram. Between the two different models, the three input appears to be able to predict a wider range of values, but both models have a tendency to predict larger dust values than \textit{Bagpipes}. 

On the right, we take the difference of the {\it Bagpipes} and predicted values, as this helps us visualise how close the individual predictions are. The shape of the distribution of both are similar, with a slight skew towards the left, indicating an over-prediction in values compared to the SED fitting. Overall, the MAE from the one input is $=0.26$, a touch larger than that of the three input $=0.24$. The standard deviation and variance from the one input is also slightly larger. This continues a similar trend that also appears in test results, where the predictions from the models skew larger and the model using three inputs is slightly better than the one input.

We also compare different methods for finding A$_{V}$ in Figure \ref{fig:bagpipes} using {\it Bagpipes} and the Meurer relation \citep{Meurer1999}. This relationship between dust extinction and the UV-continuum slope $\beta$ is as follows :
\begin{equation}
    A_{IRX-\beta} = 4.43 + 1.99 \beta
\end{equation}
At \textit{z} $\sim$ 0, this relationship has been shown to be a good predictor of dust extinction in galaxies, but does not always generalise to high-redshift galaxies \citep{Koprowski2020, Liang2021}. 

This is a common way to determine the value of $A_{V}$ based on the UV slope, and it is a method which is independent of the SED fitting discussed earlier.  The way we carry the measurement of $\beta$ out is explained in detail in \citet{Duan2023}, whereby we measure the $\beta$ slope for our galaxies through direct measurements of the slope of each NIRSpec spectrum within the UV range by fitting a power-law and measuring its slope. As we are using the actual spectrum for this measurement, we are directly measuring this slope and not inferring it from photometry.  The comparison of this 'Meurer' method with the {\it Bagpipes} measurements is shown in Figure \ref{fig:bagpipes}.  

This comparison between the {\it Bagpipes} and Meurer values also shows large differences and a larger standard deviation as compared to that of our CNN machine predictions, $\sigma$ = 1.52 to 0.34 (Figure \ref{fig:jwstav}). The range in the differences is also smaller from our machine predictions, meaning that the SED fitting method and our machine learning method agree better than the Meurer relation and SED fitting.  This  shows that our machine is at least comparable in accuracy with those based on the Meurer relation, which is commonly used to find the value of $A_{V}$.

However, what we find using all three of these analysis methods is that the dust content of distant galaxies observed with JWST is relatively low with values $A_{V} < 0.6$ and on average around $A_{V} \sim 0.3$. This means that distant galaxies, those with higher star formation rates especially, which are the ones that have successful spectroscopy, are low in dust content. This is opposed to lower redshifts where dusty galaxies can have a large impact on the galaxy population.  

As mentioned earlier in this section, the other value that we compared is sSFR. Figure \ref{fig:jwstssfr} illustrates the results we obtained for sSFR using our models on the JWST samples. Again, the leftmost scatterplot shows the difference between using 10 and 100 Myr timescale, where the sSFR values from the 100 Myr are smaller. This scatterplot shows the predicted sSFR values on the \textit{x}-axis and on the y-axis the {\it Bagpipes} values.  On this plot are shown the range of values of the 100 Myr star formation scale as red points, which are much smaller than the blue points showing the 10 Myr scale. The values of sSFR predicted by our model correlates closer to that of the 100 Myr timescale {\it Bagpipes}, which indicates that the machine is predicting reasonable sSFR values. In the middle panel we compare the values of the 100 Myr timescale {\it Bagpipes} sSFR and predicted sSFR, with predictions from the one input model shown on the top and three input model on the bottom. Both models are not very accurate and are not able to predict the full range of the sSFR values from {\it Bagpipes}. Furthermore, the shape of the predictions shows a sharp peak which is quite different to that of the {\it Bagpipes} values. On the rightmost panel we examine the differences between the sSFR values from our models and {\it Bagpipes}. There is a large range and the distribution is skewed towards the right, where the model's over-predict the values. For sSFR, the results of the one input model is similar to that of the three input model, with a similar MAE (0.18 to 0.19) and standard deviation (0.20 to 0.21). This reflects the testing results on simulated data, in which the three input model performs slightly worse but neither are able to predict extreme sSFR values on both ends. 

\section{Discussion}
\label{sec:disc}

We discuss our results both in terms of the use of machine learning for measuring dust properties of distant galaxies, as well as the results that we find when we apply the models to JWST imaging data. Both of these issues are very important in extragalactic astronomy, as dust is a key component of galaxies and the formation of stars and is a significant aspect in the observability of galaxies at different wavelengths. In addition, dust evolution in galaxies is still a major mystery and finding new ways to measure it is important for a full consensus of galaxy evolution.     

We have shown that based on stimulated imaging data that we can retrieve with a high accuracy the dust content of galaxies as measured through the $A_V$ values. Results from our testing show that using only one input (F277W) and using three inputs (F277W, F356W, F444W) yield similar results for estimating all of the properties we examine. This was slightly unexpected, as it might seem that the 'colour' information provided by the multi-bands would give some indication of the properties of dust and star formation. Properties such as colour differences and dust extinction occur because dust absorbs shorter wavelengths more efficiently, so these properties are best extracted with inputs of different wavelengths. SFR has also been strongly correlated with SED shape, although we see little difference here between the one band and the three band imaging when inputted into our machine.   

However, one of the strengths of CNNs is its ability of feature extraction in images, which we utilize by looking at the morphology of galaxies. Thus a major conclusion is that even in a single band it is possible to understand aspects of the dust content of galaxies, as well as of star formation. This has already been shown in the case of SFR for measurements using single bands \citep[e.g.,][]{Bisigello2023}. As we focus in this paper on the difference in morphology between the different bands in the 3-band input, the similar predictions indicate that the morphological differences at different wavelengths do not provide substantial improvements in helping the model predict dust and sSFR values. We find that the three input model performs marginally better overall, but the small dataset for the JWST data used here makes it difficult to assess the accuracy and generalise between the one and three input models on actual data. 

A difficulty we face in taking models trained from the simulated data and then applied to the observational data from JWST is what is called dataset shift. Dataset shift occurs when the training and testing distribution are fundamentally different \citep{Moreno2012}. This can make it difficult for the model to generalize its learning from training data and applying the model to testing data, yielding poor results. Here we find a notable possible dataset shift between the training data and the set of {\it Bagpipes} values that we use. The data simulated from Illustris uses an instantaneous star formation timescale of 100 Myr \citep{Pillepich2018}. On the other hand, we had two sets of testing data from the JWST images, calculated from \textit{Bagpipes}, using a star formation timescale of 10 and 100 Myr, as well as using the Meurer relation.  We find first that through any method the dust content of galaxies at $z \sim 6-7$ are relatively low, within the lower regime of our training set, and that there is within these values considerable scatter is present.  

The leftmost plot in Figure \ref{fig:jwstssfr} shows how the two sets of \textit{Bagpipes} values correlate with our predictions, trained on a timescale of 100 Myr. The range of sSFR values between those predicted by our model is also close to the range predicted by the 100 Myr SFR-timescale values from \textit{Bagpipes}, showing how important the training data is in getting accurate predictions. The predictions our models made on the JWST data also reflects the range of sSFR values of the simulated galaxies. Extreme values from \textit{Bagpipes} that were not represented during training makes it more difficult for the neural network to predict.  This is one reason why sSFR is difficult to measure in machine learning as the values of real galaxies have longer tails in this value's distribution at low and high values than other properties such as the dust extinction.

The effects of the 10 Myr and 100 Myr SFR-timescale dataset shift is not as obvious with dust extinction, shown in the leftmost scatterplot of Figure \ref{fig:jwstav}. There is a small difference in the values of A$_{V}$ from \textit{Bagpipes} using 10 and 100 Myr SFR-timescale. However, the range of values used in training from Illustris for dust extinction was much larger, up to A$_{V} > $ 2. When we tested the models with JWST data, the models were able to retrieve small values of dust extinction. 

The uncertainty of dust measurements at the high-redshift regime also makes it difficult to determine how accurate our model is. Our predictions estimate that there is some amount of dust attenuation on these early JWST galaxies, but not as high as at lower redshifts. The simulated data that we trained the CNN models on had a wide range of dust extinction values ($A_{V} > 2$), but groups of high-redshift galaxies observed by JWST have been shown to be extremely blue \citep{Adams2022, Atek2022, Castellano2022, Finkelstein2022, Naidu2022, Ferrara2023} with very low amounts of dust attenuation (down to $A_{V}$ < 0.02) \citep{Furtak2022}. This signifies a significant difference between the properties of these high redshift JWST galaxies and the galaxy samples we trained the CNN models with.   It is a sign that dust has not had time to build up within galaxies at the time that we are observing them at $z \sim 6-7$.

The results presented here utilise a CNN machine learning tools with a structure of three convolution and four dense layers. However, other popular CNN models (i.e. LeNet, AlexNet, VGG-16) utilise far more convolution layers and larger number of variables, such as the kernel size or number of filters. During the tuning process, our models converged to the maximum number of filters we set. This may indicate that more filters could help better capture the features of the galaxies. However, more hyperparameters would lead to a significant increase in parameter space, as each successive layer builds upon the number of outputs of the previous layer. With longer training time and increased computing power, the structure could be further tuned for a better performing model.

%The understanding of galaxy properties in the early universe is one of the main goals of JWST.  Having a way and method for carrying this out using various techniques is important for understanding how galaxy evolution and formation has occurred over cosmic time.  One of the main ways we can do this is through using machine learning to understand the properties of dust within galaxies.  Our approach shows that within the limitations of an advanced simulation, we are able to reproduce the values.  

Future work in this direction could include incorporating MIRI images into the training dataset when available.  This will not work for all galaxies as MIRI detections of such high redshift galaxies will be difficult except when examining very long exposure times \citep[e.g.,][]{Li2023}.  The longer wavelength capabilities of deep MIRI data might be able to observe even higher redshift galaxies as thus find and locate spectral properties inaccessible in the NIR wavelengths.  Studies such as \citet{Vika2015, Mager2018} have shown that morphology has a strong wavelength dependence, so MIRI images may be able to contribute more features to improve the prediction accuracy of the machine learning method.  Ultimately, if we can find a direct approach to dust extinction we can train samples on large collections of galaxies with these measurements which may be a superior method than using simulated imaging.

\section{Summary}

We present in this paper a Machine Learning analysis of simulated galaxy images from Illustris TNG to retrieve the physical properties of dust extinction and specific star formation rate in high-redshift galaxies, purely from JWST imaging. Our results show that the CNN model we construct and calibrate can reproduce these values well within the simulated images. 

For this analysis we use simulated images from Illustris created as imaged in the JWST  bands and within NIRCam imaging specification, including the commonly observed F277W, F356W and F444W filters. We create a set of models using only the F277W band as an input, and another model using all three of these bandpass images as inputs. We find that the three input model perform slighlty better across $E(B-V)$ and $A_{V}$, but performs slightly worse at predicting sSFR. We use MAE and MSE values to judge how close the predictions are to the real values - for the colour excess E(B-V), the MAE value of the one input model is 1.36 $\times$ 10$^{-2}$ which is only slightly larger than that of the three input, giving a MAE of 1.31 $\times$ 10$^{-2}$. The same trend whereby the MAE is slightly smaller than MSE for the three input model is seen in the retrieved dust extinction A$_{V}$ and R-values.  Using standard deviations and means, we find again that the three band model appears to be more robust and performs slightly better against the one input model in estimating colour excess, dust extinction and R-value. 

We compare our machine learning results to fits from SED fitting models using \textit{BagPipes} where the input is the entire JWST multi-wavelength data to determine how consistent these approaches are. The one input model predictions of sSFR agrees slightly better with the \textit{BagPipes} values compared to the predictions of the three input model. The MAE of the one input model is 4.56 $\times$ 10$^{-2}$ compared to the MAE of the three input model, 6.88 $\times$ 10$^{-2}$. The standard deviation of one input is also smaller than the three input, with a value of 6.71 $\times$ 10$^{-2}$. As effects of colour excess and dust extinction are calculated from the differences between different bandwidths of images, it is reasonable that the three input model is more robust and performs better than the one input model. However, we observe similar values and the error from both machine learning runs.

Data taken from JWST allows us to test our machine learning methods on observed galaxy images. Using a sample of high redshift galaxies from the JADES and CEERS data release previously analysed in \cite{Duan2023} which have spectroscopic redshifts, we compare our predictions against values estimated by SED fitting methods from \textit{Bagpipes}.  Thus, we are able in this paper to compare dust measures from both the traditional SED fitting technique and those based on our new formalism.  In general, we find similar trends as from training when comparing between these two.  For dust extinction, the MAE of one input predictions is 0.26, slightly larger than the MAE of three input predictions, which is $\sim 0.24$. In values of sSFR, the one input has a slightly smaller MAE, 0.18 against 0.19, the MAE of the three input model. However, we do not know which of the measurements, either the SED fitting or the ML derived values, are 'correct'. We do find that both methods give low dust extinction values for distant galaxies, thus suggesting that dust has not built up to a significant amount in the earliest galaxies (see also Austin et al. 2024, in prep).
 
We demonstrate this general problem of measuring measure of dust content from e.g., A$_{\rm V}$ values by comparing the results using the $\beta$ method from the Meurer relation. Compared to the values from the Meurer relationship, our ML predictions have a similar average error, but a smaller standard deviation (SD$_{Meurer}$ = 0.89 against SD$_{CNN}$ = 0.36). The errors show that our new machine learning method is comparable to other, more traditional, methods used to determine dust attenuation. 

In conclusion, we find that the CNN model is able to predict dust values well, with a smaller than expected difference between the performance using one and three bandwidth images. Applying the models to JWST galaxies showed some potential limitations with the predictions of CNN models compared to SED fitting. However, what is clear is that this method can be applied in bulk successfully to derive a measure, within some uncertainty, compared with full SED fitting methods. 

\section*{Acknowledgements}

We acknowledge support from the ERC Advanced Investigator Grant EPOCHS (788113), as well as  studentships from STFC. LF acknowledges financial support from Coordenação de Aperfeiçoamento de Pessoal de Nível Superior - Brazil (CAPES) in the form of a PhD studentship. This work is based on observations made with the NASA/ESA \textit{Hubble Space Telescope} (HST) and NASA/ESA/CSA \textit{James Webb Space Telescope} (JWST) obtained from the \texttt{Mikulski Archive for Space Telescopes} (\texttt{MAST}) at the \textit{Space Telescope Science Institute} (STScI), which is operated by the Association of Universities for Research in Astronomy, Inc., under NASA contract NAS 5-03127 for JWST, and NAS 5–26555 for HST.

%%%%%%%%%%%%%%%%%%%%%%%%%%%%%%%%%%%%%%%%%%%%%%%%%%
\section*{Data Availability}

The inclusion of a Data Availability Statement is a requirement for articles published in MNRAS. Data Availability Statements provide a standardised format for readers to understand the availability of data underlying the research results described in the article. The statement may refer to original data generated in the course of the study or to third-party data analysed in the article. The statement should describe and provide means of access, where possible, by linking to the data or providing the required accession numbers for the relevant databases or DOIs.



\bibliographystyle{mnras}
\bibliography{template} % if your bibtex file is called example.bib

%\begin{thebibliography}{99}
%\end{thebibliography}

%%%%%%%%%%%%%%%%%%%%%%%%%%%%%%%%%%%%%%%%%%%%%%%%%%

%%%%%%%%%%%%%%%%% APPENDICES %%%%%%%%%%%%%%%%%%%%%

%\appendix


%%%%%%%%%%%%%%%%%%%%%%%%%%%%%%%%%%%%%%%%%%%%%%%%%%

% Don't change these lines
\bsp	% typesetting comment
\label{lastpage}
\end{document}